# A new job migration algorithm to improve data center efficiency

**Federico Calzolari**[1]

*Scuola Normale Superiore*
*Piazza Cavalieri 7, 56126 Pisa PI, ITALY*
*E-mail:* `federico.calzolari@sns.it`

**Silvia Volpe**

*University of Pisa, Information Engineering Department*
*Via G. Caruso 16, 56122 Pisa PI, ITALY*

The under exploitation of the available resources risks to be one of the main problems for a computing center. The growing demand of computational power necessarily entails more complex approaches in the management of the computing resources, with particular attention to the batch queue system scheduler. In a heterogeneous batch queue system, available for both serial single core processes and parallel multi core jobs, it may happen that one or more computational nodes composing the cluster are not fully occupied, running a number of jobs lower than their actual capability. A typical case is represented by more single core jobs running each one over a different multi core server, while more parallel jobs - requiring all the available cores of a host - are queued. A job rearrangement executed at runtime is able to free extra resources, in order to host new processes. We present an efficient method to improve the computing resources exploitation.

[1] Speaker

## 1. Introduction

The increase of the computational power leads necessarily to more complex approaches in the resources exploitation and management. One of the main problems nowadays for a computing center is given by the under exploitation of the available resources.

It may happen that in a heterogeneous batch queue system, available for both serial single core processes and parallel multi core jobs, one or more computational nodes of the cluster are serving a number of jobs lower than their actual capability. A typical case is represented by more single core jobs running each one over a multi core server, while more parallel jobs - requiring all the available cores of a host - are queued.

We present an efficient method to improve the computing resources exploitation.

## 2. Aims

The aims of our project are to improve the farm exploitation in terms of running jobs, and increase the computing farm efficiency, by reducing the free job slots.

## 3. The problem

Given a computing farm with multicore servers with a batch queue system on board, it may enter a blocking situation, where even if many servers have some free job slots (in green in Figure 1) the queued jobs are retained.

It may occur for example if the first queued job requires more job slots than the available ones on a single server.

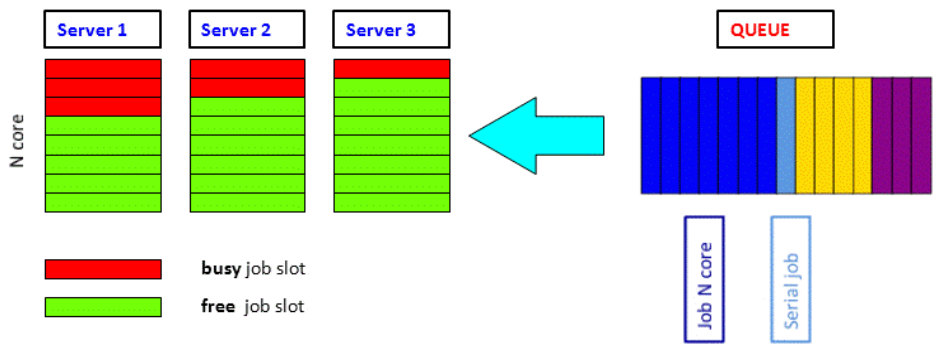

Figure 1: The blocking condition

### 3.1.1 The batch queue system

The issues related to the computing farm under-exploitation depend on the batch queue system behavior [1] [2]:

- the batch queue system cannot modify the queued jobs order;
- the scheduler has to respect fairshare and job priorities;
- the batch queue system cannot move jobs at runtime.

## 4. The possible solutions

The classical and probably most used solution is the cluster partition: the computing farm is divided into independent blocks, in order to run in separate environments and over separate sets of nodes the serial jobs (requiring a single core) and the parallel jobs (multi-core).

This solution entails the loss of all the benefits coming from the sharing of the computing resources, and the under-exploitation of the computing farm if one of the two partition is not fully used.

In order to solve this problem, we suggest a new strategy: a Job mover able to displace and rearrange the jobs over the farm at runtime.

### 4.1.1 The problem complexity

The problem of the possible permutations achieved by moving a set of jobs, each one requiring a variable number of cores, over a set of servers, is described by an NP-complete complexity class.

Due to the difficulty in finding the best solution, we focused on searching for a solution able to improve the current load status of the cluster, certainly not the optimum one.

## 5. The Job mover

The idea is to pile up the maximum number of jobs over the minimum number of hosts, compatibly with the available CPU and memory on the single hosts, in order to fill as many contiguous job slots as possible, as shown in Figure 2.

In this way the running jobs do not suffer any performance loss, and at the same time the farm may gain some contiguous job slots able to host new parallel multi-core jobs otherwise blocked in queue.

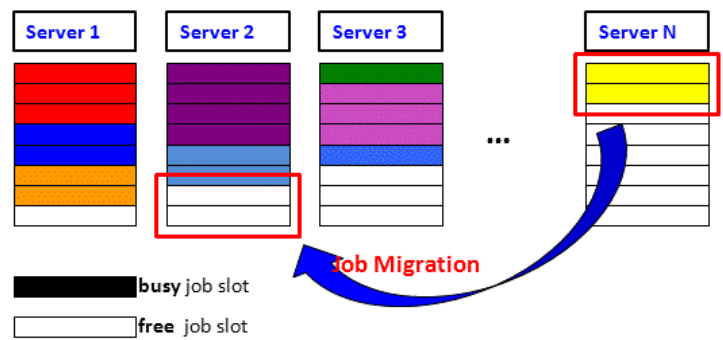

Figure 2: Job Migration

A job displacement, executed at runtime in order to stack up more processes over a single multi core server, is able to free extra resources able to host new processes - both single or multi core.

In order to ensure a full hosts exploitation, and prevent at the same time the overload of one or more nodes in the cluster, the job migration takes place only under certain conditions.

We also paid special attention to avoid a too frequent job displacement, damaging the global performances.

### 5.1.1 The simulator

A prototype of job mover [3] has been developed with the aim of freeing the best part of otherwise unavailable resources in a computing cluster. We started implementing a batch system and queue simulator, in order to test the efficiency of several job rearrangement algorithms.

We just developed a software to simulate the behavior of the farm, the queue, the job mover algorithm, and collected the statistics with and without the applied patch.

Defining an exploitation parameter [4], strictly connected to the cluster load, we implemented two algorithms able to increase the computational resources load - limited for each server by the number of available cores.

The cluster and queue simulator may also be used to test other rearrangement algorithms, in order to achieve an even better result in the cluster exploitation.

### 5.1.2 Requirements

The first request is to set up the batch system behavior in order to fill the minimum number of servers, instead of balance the load between all the available servers. In this way the queued jobs will try to fill the job slots available in the server number N before they start filling the server number N+1.

To take advantage of the features provided by the job migration system, the only requirement is the job checkpoint capability - a complete disk and memory dump is needed for a job freeze and its immediately subsequent restart on another host. This capability is today guaranteed from the major batch queue systems available: PBS, LSF, SGE. In other terms, the batch queue system needs to provide the job migration feature.

The second step is to rearrange the jobs allocation at runtime once per hour, considering the free resources available in the farm.

The requirements in terms of CPU, RAM, disk and I/O for a single job need to be compliant to the given acceptance scheme: the farm is able to accept only N core jobs, requiring each one a fraction (N/C) of RAM, disk, and I/O bandwidth; where N is the number of required cores, and C the single server cores number.

## 6. Tests and results

### 6.1.1 Use cases

The use cases we analyzed are characterized by mixed serial (mono-core) and parallel (multi-core) jobs, where parallel jobs are spread between 2 and the maximum number of cores available on a single server; the job running time is included between 1 hour and 15 days; the queued jobs distribution is random or sequential. We collected one year of simulated data.

The following graphs display the overall CPU time used with respect to the time - both of them expressed in hours.

### 6.1.2 Random jobs

The first analyzed use case is set up by a random distribution of jobs requiring 1 to 8 cores, with a life time between 1 hour and 15 days, and the queue filled in a random way. The farm is composed by 128 servers, 8 cores each one. The data acquisition covers one year.

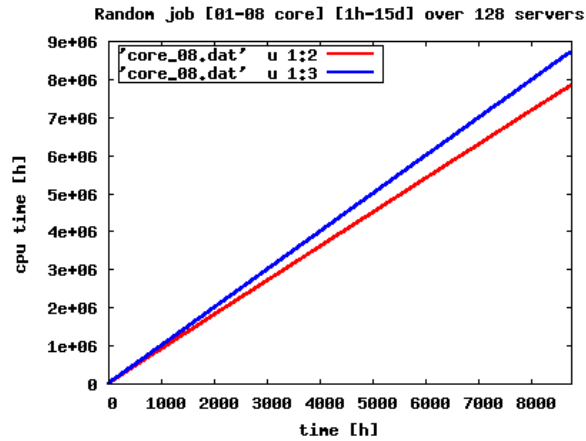

Figure 3: Random jobs

In Figure 3, in red is shown the natural evolution - without job mover patch; in blue the modified evolution - with the job migration system applied.

The efficiency improvement is about 12%, with an amount of moved jobs of 4717, over a total of 10726 jobs run on the farm in one year (43% of jobs are moved during their lifetime).

### 6.1.3 Worst case condition

The second analyzed use case is set up by a repeated sequence of serial mono-core long term jobs, followed by parallel full-core short term jobs. The farm is composed by 10 servers, 8 cores each one. The data acquisition covers one year.

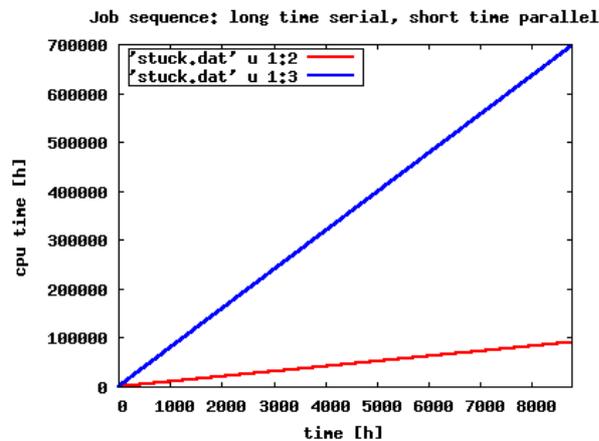

Figure 4: Blocking condition

In Figure 4, in red is shown the natural evolution - without job mover patch; in blue the modified evolution - with the job migration system applied.

The efficiency improvement is about 800% (8 times), with an amount of moved jobs of 2239, over a total of 17156 jobs run on the farm in one year (13% of jobs are moved during their lifetime).

Even if the situation is made ad hoc, it is more frequent than expected.

## 7. Algorithm efficiency

### 7.1.1 Core number per server

In Figure 5 is shown the algorithm efficiency with respect to the number of cores (8, 12, 24, 48) per server, in a 128 servers farm, with completely random jobs sequence and one year of data acquisition. The efficiency improvement is between 11 and 13%.

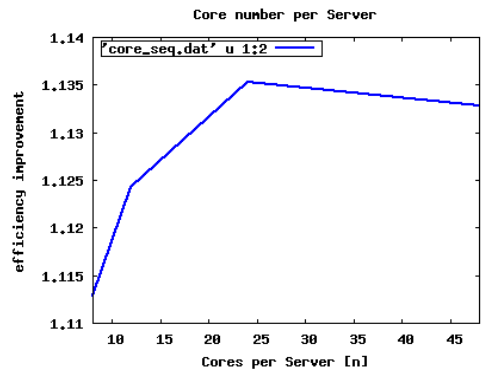

Figure 5: Algorithm efficiency vs Core number per server

### 7.1.2 Server number per farm

In Figure 6 is shown the algorithm efficiency with respect to the number of servers (3, 5, 10, 20, 50, 100) per farm, with 8 cores servers, with completely random jobs sequence and one year of data acquisition.

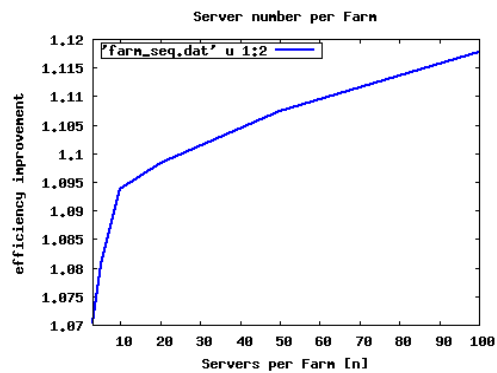

Figure 6: Algorithm efficiency vs Server number per farm

The efficiency improvement is between 7 and 12%, with a percentage of moved jobs varying between 10 and 50% depending on the farm size and jobs type. The farm efficiency increases with the increasing of the server number.

## 8. Green computing

A secondary effect, probably not less appealing by the point of view of the "green computing", is represented by the power efficiency improvement through a dynamic job rearrangement, with an energy saving up to 90% in some particular cases - more frequent than expected.

By the use of a remote controlled power supply, it is possible to switch off the unused hosts, waiting to be switched on upon request.

## 9. Conclusions

The system, developed at Scuola Normale Superiore in collaboration with the Information Engineering Department of the University of Pisa [Italy], is able to provide an increase in the number of running jobs over a general purpose computing cluster.

The computing power efficiency improvement, simulated over a large computing farm, varies from 7 to 13% in typical use cases, reaching 800% in some particular situations, with a very few constraints in terms of job requirements, and no modifications required neither to the batch queue system, nor to the scheduler.